\documentclass[twocolumn,prb,amsmath,amssymb,amsfonts,superscriptaddress,floatfix,aps,showpacs,notitlepage]{revtex4-1}

\usepackage{graphicx}
\usepackage{epstopdf}
\usepackage{dcolumn}
\usepackage{bm}
\usepackage{rotating} 
\usepackage{color}
\usepackage{subfigure}
\usepackage{natbib}
\usepackage{enumitem}
\usepackage[Symbol]{upgreek}
\usepackage[symbol*]{footmisc}
\usepackage{lipsum}
\usepackage{amsmath}                    
\usepackage{amssymb}                    

\usepackage{textcomp}					

\usepackage[version=3]{mhchem}
\usepackage{units}
\usepackage{verbatim}
\usepackage{url}
\usepackage[utf8]{inputenc}
\usepackage[T1]{fontenc}

\newcommand{\ket}[1]{\mbox{$| #1 \rangle$}}

\begin{document}
	
	\title{Kramers degeneracy and relaxation in vanadium, niobium and tantalum clusters}

	\newcommand{\imm}{Radboud University, Institute for Molecules and
		Materials, Heyendaalseweg 135, 6525 AJ Nijmegen, The Netherlands}
	
	\newcommand{\felix}{Radboud University, Institute for Molecules and
		Materials, FELIX Laboratory, Toernooiveld 7c, 6525 ED Nijmegen,
		The Netherlands}
	
	\author{A. Diaz-Bachs}
	\affiliation{\imm} 
	
	\author{M. I. Katsnelson}
	\affiliation{\imm}
	
	\author{A. Kirilyuk}
	\email{a.kirilyuk@science.ru.nl}
	\affiliation{\imm}
	\affiliation{\felix}

	\date{\today}
	
	\begin{abstract}
		In this work we use magnetic deflection of V, Nb, and Ta atomic clusters to measure their magnetic moments. While only a few of the clusters show weak magnetism, all odd-numbered clusters deflect due to the presence of a single unpaired electron. Surprisingly, for majority of V and Nb clusters an atomic-like behavior is found, which is a direct indication of the absence of spin-lattice interaction. This is in agreement with Kramers degeneracy theorem for systems with a half-integer spin. This purely quantum phenomenon is surprisingly observed for large systems of more than 20 atoms, and also indicates various quantum relaxation processes, via Raman two-phonon and Orbach high-spin mechanisms. In heavier, Ta clusters, the relaxation is always present, probably due to larger masses and thus lower phonon energies, as well as increased spin-orbit coupling.
	\end{abstract}
	
	\maketitle
	
	\section{Introduction}
	Kramers degeneracy theorem \cite{Kramers1930}  states that every energy eigenstate of a time-reversal symmetric system with non-integer total spin is at least doubly degenerated. The basis states of the system are Kramers-conjugate, i.e. they are related to each other by the time-reversal operator. The immediate consequence of this is that for such a system, spin-lattice coupling is prohibited, because any spin-phonon operator is invariant under time reversal, and therefore has zero matrix elements for the transitions between such states. This selection rule, also known as the Van Vleck cancellation\cite{VanVleck1940}, implies that the lattice excitations cannot be responsible for the relaxation between two Kramers conjugated states. Therefore, S$_{z}$ is a good quantum number similar to that of an isolated atom.
	
	The relaxation can nevertheless happen via either Raman or Orbach mechanisms\cite{Orbach1961}. The first one includes excited vibrational states and is thus temperature dependent. The second type of relaxation involves excited spin states and is active in systems with the total spin larger than 1/2. Moreover, there are other, more material-dependent mechanisms of the relaxation, such as for example the electronuclear spin entanglement\cite{Ruiz2014}.
	
	In the absence of the relaxation, a system with the minimum half-integer spin may represent the smallest possible magnetic bits thus creating a new paradigm in magnetic storage technology. The understanding of the exact behavior of the relaxation mechanisms is therefore very important, both for the possible applications and for the fundamental understanding of the quantum decoherence processes. The majority of studies are focused on the behavior of single magnetic ions\cite{Piligkos2015}. In larger system, the decoherence processes are usually considered too strong for any realistic appearance of the spin blocking. This is particularly true when a non-isolated system is considered. 
	
	Gas-phase atomic clusters represent ideal model systems\cite{deHeer1993,Reinhard2003}, used to understand various phenomena in totally different areas of science, from nuclear physics to crystal growth. The condensed matter properties such as magnetism\cite{Alonso2000,Billas1994}, are combined with the molecular reproducibility of their structure. All energy levels in the clusters are discrete and tunable by simply varying their size, leading to the unique possibility to tune the microscopic correlations and from this the macroscopic properties, to our needs.
	
	Here we demonstrate that the Kramers degeneracy leads to the spin blocking in small gas-phase clusters of early d-metals, such as vanadium and niobium. In such clusters, the interaction with external bath are fully excluded. Due to the odd number of electrons per atom, there is always a non-zero total spin in the clusters with odd number of atoms. Several of the clusters with the total spin 1/2 showed the blocked-spin behavior on the time scale of the experiment ($\sim$0.1~ms), in spite of the highly populated rotational states. Clusters with larger magnetic moments, though also corresponding to the half-integer spin, showed the clear superparamagnetic behavior, indicating the Orbach relaxation mechanism. Moreover, introducing vibrational excitation in a cluster also leads to the appearance of relaxation via Raman mechanism.

	\section{Experimental setup}
	The setup used for these series of experiments consists mainly of three parts: the cluster source, the deflection magnet, and the position sensitive time of flight mass spectrometer (PSTOFMS). The clusters are produced in the source chamber by laser ablation of a metal rod. The source is of a Milani-de Heer kind \cite{deHeer1990}. The laser is a Nd:YAG laser with frequency doubling to produce 532~nm light. It is focused on the rod placed inside a cavity of a tuneable volume. A pre-cooled He gas is injected into the cavity, resulting in cluster nucleation and growth. Subsequent cooling is taking place in an extension tube of 20~mm length. Once the clusters left the source through the nozzle, the cluster beam is skimmed through a conical skimmer of 1~mm diameter. The beam is further formed by an adjustable slit 1~m downstream. For the experiments described here, the size of the slit was 2$\times$0.5~mm$^2$. 
	
	The source is mounted on a closed-cycle cryocooler and can be cooled down to 20~K. Before entering the cluster source, the carrier gas passes through a copper tube wound around the cryocooler head. With this, the minimum estimated temperature of the cluster ensemble was about 40~K. 
	
	Before the deflection magnet, cluster pass a mechanical chopper, that serves to select the cluster bunch with a defined velocity. The velocity is thus defined by the time difference between the chopper and the ionizing excimer laser 1.68~m further downstream. Depending on the temperature of the cluster source, the velocity varied between 400 and 700 m/s. In fact, the measured velocity always corresponded to a higher temperature than that of the source, by roughly 20~K. In the following, the temperature of the source is indicated.  
	
	The deflection magnet is of the Rabi-type two-wire magnet \cite{Rabi1934} which produces an inhomogeneous magnetic field, able to reach 2.4~T with up to 650~T/m gradient. The gradient is calibrated using atomic beam of aluminium; note however that atomic deflection can only calibrate the gradient and not the field itself. The latter was determined by a Hall-probe measurements, as well as via the agreement with the calculated values. 
	
	After the magnet there is a flight distance of about 1~meter after which the clusters enter the PSTOFMS. An excimer laser at the wavelength of 193~nm is used to ionize the clusters. The mass-spectrometer is somewhat detuned from the ideal spatial focusing conditions, that sacrifices a little part of mass resolution for the convenience of position sensitivity. The clusters are detected with the help of a microchannel plate (MCP). PSTOFMS is calibrated by scanning a narrowly focused excimer laser beam across the cluster beam. As an example, at the used electrode settings, the position sensitivity for the cluster Nb$_{20}$ is 19.2~ns/mm. This sensitivity obviously scales as the square root of mass\cite{Milani1991}.
	
	\section{Results}
	\subsection{Vanadium}
	Vanadium is a 3d metal, with an electronic configuration [Ar]3$d^3$4$s^2$. From the early research, vanadium clusters were expected to show magnetic properties, due to their 3d electrons. As a result, a large number of theoretical studies have been carried out~\cite{Liu1991,Dorantes1993,Lee1993,Ray1999,Zhao1995,Alvarado1994}. In contrast, the experiments are scarce~\cite{Douglass1991}. The majority of these works disagree with each other, though. Liu et al., found a magnetic moment of 2.89~$\mu_B$ for V$_{9}$ while no magnetic moment at all for V$_{15}$. The same work also estimates that a chain between 2 and 7 vanadium atoms would achive a magnetic moment around 4~$\mu_B$ per atom~\cite{Liu1991}. Dorantes-D{\'a}vila et al.\ studied the magnetic moments as function of the parameter J/W, where J is the exchange integral and W is the bulk bandwidth; they found a total magnetic moment between of 0-4~$\mu_B$ for ferromagnetic V$_{9}$ cluster \cite{Dorantes1993}, while the value decreased to 0-3~$\mu_B$ for an antiferromagnetic order. V$_{15}$ showed a magnetic moment of 0-4~$\mu_B$. Lee et al., found different values for the magnetic moment of V$_{9}$ depending on the lattice parameter~\cite{Lee1993}, going from 0.33-2.78~$\mu_B$ to higher values as the lattice parameter becomes larger, and the magnetic moment of only 0.07~$\mu_B$ for V$_{15}$. Zhao  et al.\ studied vanadium clusters between V$_{2}$ and V$_{15}$, with magnetic moments only showing the values of 1~$\mu_B$ for V$_{9}$ and 0.6~$\mu_B$ for V$_{15}$ \cite{Zhao1995}. Wu et al.\ calculated the magnetic moments for the two lowest energy configurations, finding a magnetic moment of 3~$\mu_B$ for V$_{3}$ for one configuration and 5~$\mu_B$ for the other, 0~$\mu_B$  was determined for both V$_{4}$, 1 and 3~$\mu_B$ for V$_{5}$, 2 and 0~$\mu_B$ for V$_{6}$, 0.98 and 3.1~$\mu_B$ for V$_{7}$, 2~$\mu_B$ for V$_{8}$ and 3 and 0.99~$\mu_B$ for V$_{9}$~\cite{Ray1999}.
	
	On the experimental side, Douglass et al.\ failed to observe any sizable magnetic deflections of vanadium clusters \cite{Douglass1991}.  According to the resolution of their experiment they were able to calculate a maximum value for the magnetic moment per atom for of 0.59~$\mu_B$ for V$_{9}$  and 0.18~$\mu_B$ for V$_{99}$.
	
\begin{figure}[tb]
		\centering
		\includegraphics[width=0.9\linewidth]{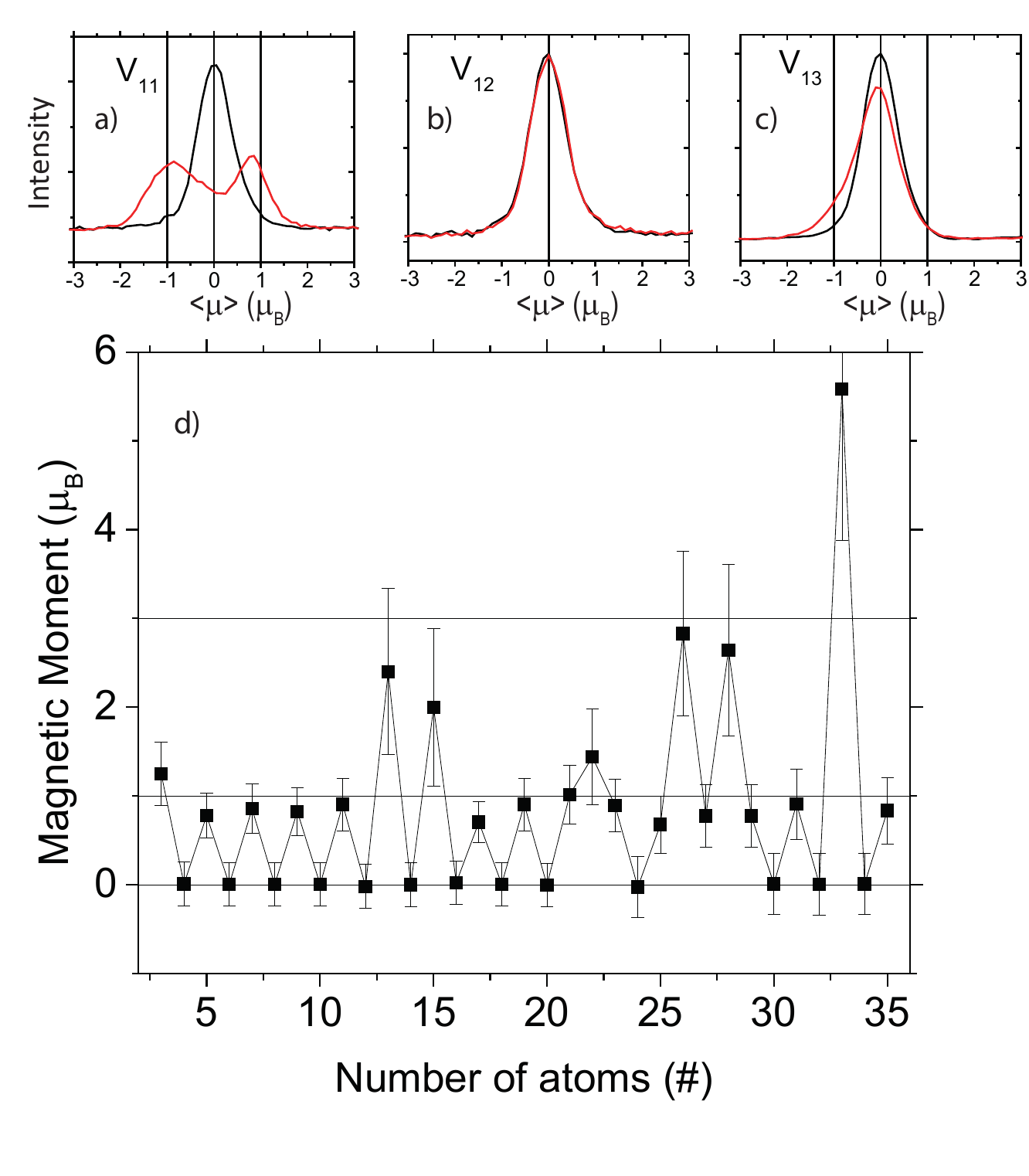}
		\caption{Top: Different kind of behaviors of vanadium clusters: a) V$_{11}$, which shows an atomic-like splitting in two states; b) V$_{12}$, which is not magnetic; c) V$_{13}$, which is a superparamagnetic cluster; all of them obtained at a temperature of 25~K; black line is for zero magnetic field, while the red line corresponds to the magnetic field of 2.4~T; the dashed lines point at $\pm$1~$\mu_B$, in case that the behavior is atomic-like.  Bottom: d) Evolution of the magnetic moment as a function of cluster size.}
		\label{mag-V}
	\end{figure}

	A summary of our deflection data for vanadium clusters at the lowest used temperature of 25~K is shown in Fig.~\ref{mag-V}. The graphs in Fig.\ \ref{mag-V}(a-c) illustrate the typical different deflection profiles found for clusters of different sizes. Thus, V$_{11}$ clusters show behavior indistinguishable from that of a purely atomic beam, such as for example the Al atoms used to calibrate the magnet. On the other hand, any deflection is clearly absent in the case of V$_{12}$. A single-sided deflection is observed in the case of V$_{13}$, which is a clear signature of a cluster with a superparamagnetic behavior. In general, the clusters show a distinct odd-even behavior, see Fig.\ \ref{mag-V}(d), with even-number-of-atoms clusters showing no magnetism whatsoever, and the majority of odd-numbered clusters demonstrating an atomic-like behavior, similar to V$_{11}$. This alternation of non-magnetic behaviour for even number of atoms clusters and magnetic behaviour for odd number of atoms clusters was pointed by de Heer et al.~\onlinecite{deHeer2004} 
	
	A few clusters, such as V$_{13}$, V$_{15}$, and V$_{33}$, as well as the even-numbered exceptions V$_{22}$, V$_{26}$ and V$_{28}$, show the superparamagnetic single-sided deflection, such as expected for clusters. Their magnetic moments are thus calculated by taking into account the cluster temperature and the Brillouin paramagnetic susceptibility formula. Note that the usually applied in such cases Langevin law is not applicable as the magnetic moments are small. Nevertheless, the difference between the two is at any rate within the error bar, which is largely due to the rather approximate determination of the cluster temperature. Thus V$_{13}$ and V$_{15}$ have a magnetic moment of 2.5-3~$\mu_B$, while for V$_{33}$ magnetic moment of $\sim$5~$\mu_B$ is found. A value of about 3~$\mu_B$ is also found for V$_{26}$ and V$_{28}$, and a small deflection corresponding to $\sim$1.5~$\mu_B$ is derived for V$_{22}$. 
	
	In contrast, in the majority of odd-numbered clusters an ideal atomic-like behavior is observed, giving a rather precise value of 1~$\mu_B$ magnetic moment per cluster. Note that this value is not affected by our poor determination of the cluster temperature, nor by an error in the value of magnetic field. 
	
	To follow the further discussion, note that at the temperatures of the experiment, of the order of 25-40~K, the majority of the clusters are in the vibrational ground state, given the fact that the vibrations of the clusters starting at about 100~cm$^{-1}$.~\cite{Fielicke2004} In contrast, the rotational constant for e.g.\ V$_{11}$ can be estimated to be of the order of 0.01~cm$^{-1}$. Therefore, the rotational states are heavily populated. Given the time scale of the experiment, about 0.1~ms (i.e.\ that long the clusters travel through the magnet), such atomic-like behavior can mean one thing only: as the spin ``up'' stays ``up'' during all this time, there cannot be any interaction between the spin and the lattice. How is this possible? 
	
	The possibility is thus provided by the Kramers theorem, that specifically forbids the time-symmetric interactions in a system with a half-integer total spin. Because of this, the direct spin-lattice interaction is absent. The processes that may allow such interaction are of the higher order, such as Raman process via phonons, or Orbach relaxation via various spin states. In our case, first of all, the phonons are absent, and second, with the total spin of $S=1/2$, there are also no additional excited spin states. 
	
\begin{figure}[tb]
		\centering
		\includegraphics[width=1.0\linewidth]{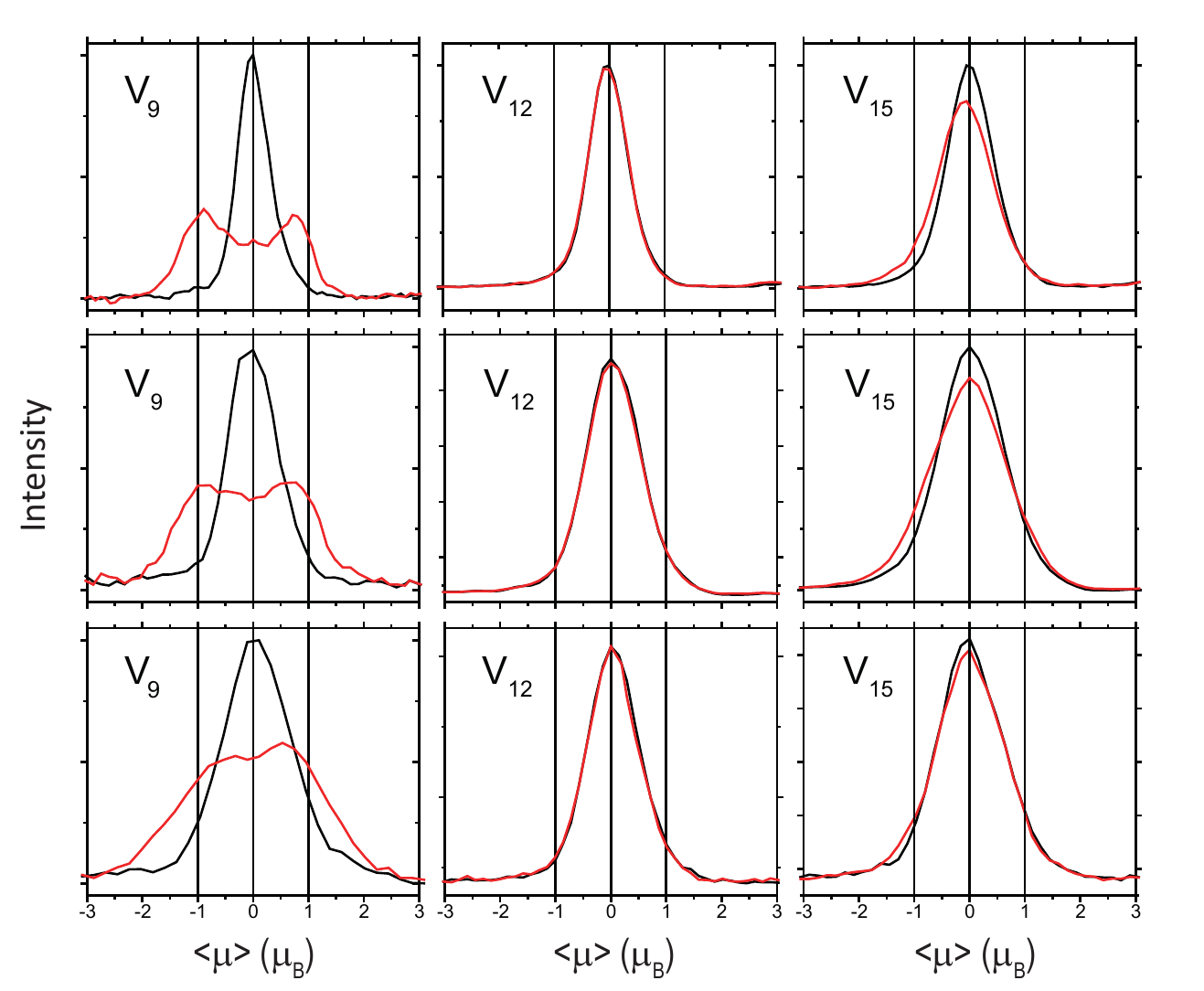}
		\caption{Evolution of deflection profile for vanadium clusters as a function of T. Top row shows deflections at 25~K. Middle row shows deflections at 40~K. Bottom row shows deflections at 60~K. Black line is for zero magnetic field, while the red line corresponds to the magnetic field of 2.4~T; the lines that indicate $\pm$1~$\mu_{B}$ and 0~$\mu_{B}$, correspond to the peaks expected for the atomic-like behaviour.}
		\label{temp-V}
	\end{figure}

	Next we have measured the temperature dependence of the deflections, shown in Fig.\ \ref{temp-V}. Unfortunately, the deflections quickly decrease with increasing the temperature, because of the simultaneously increasing cluster velocity. Nevertheless the experiments for the source temperatures of 25, 40 and 60~K all showed resolvable deflections for all the magnetic clusters mentioned before. For higher source temperature of 100~K only the lighter clusters showed deflections, V$_{3}$-V$_{11}$; heavier clusters and specifically none of the single sided deflections could be distinguished. 
	
\begin{figure}[tb]
		\centering
		\includegraphics[width=1\linewidth]{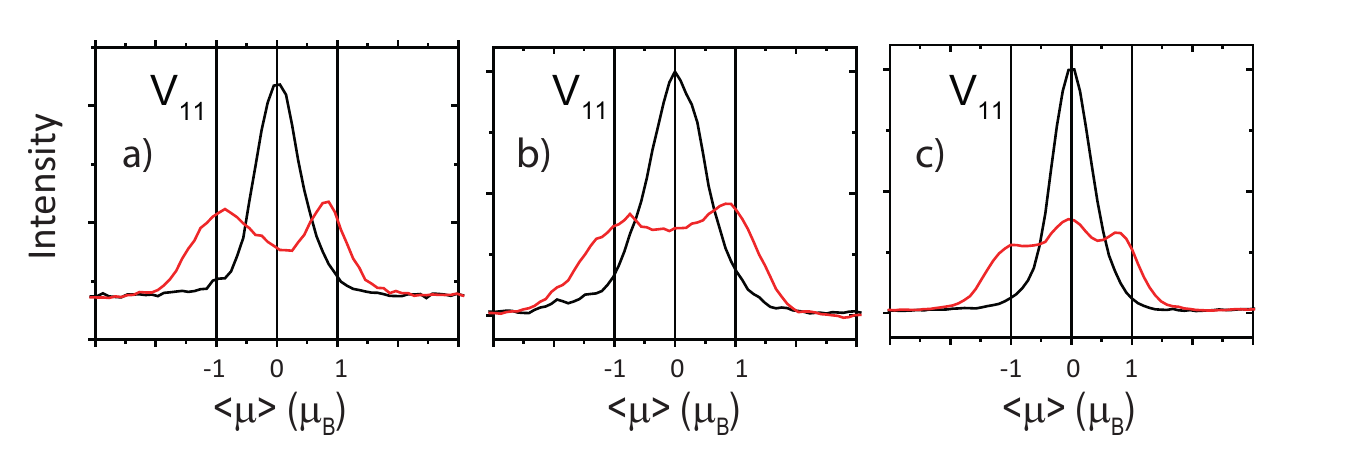}
		\caption{Relaxation for V$_{11}$ cluster. On the left panel, good thermalization can be seen, obtained with 6~bar of helium backing pressure and optimized nozzle shape. On the centre panel, 2~bar of helium pressure and optimized nozzle were used. On the right panel, 2~bar of helium pressure and different nozzle were used; the dashed lines point at $\pm$1~$\mu_B$, in case that the behaviour is atomic-like.}
		\label{pressure}
	\end{figure}

	Another way to change the thermal conditions of the clusters is to adjust the backing pressure of the He gas in the cluster source. Reducing the pressure also reduces the degree of thermalization. Alternatively, this is also to some extent affected by the shape of the nozzle. Fig.\ \ref{pressure} shows the deflection profile obtained at the source pressure of 2 bar as compared to the previously shown data at 6 bar. A central (undeflected) peak appears at 2 bar, which corresponds to a number of clusters that show a relaxation of their spin on the time scale of the experiment. This is thus an indication that the relaxation appears in non-thermalized clusters, apparently due to the residual vibrations, that allows Raman mechanism of the relaxation. Though less pronounced, a qualitatively similar effect is obtained by a different shape of the nozzle. 
	
	As mentioned above, in the only experimental work on V clusters by Douglass et al.~\cite{Douglass1991}, no sizable magnetic deflection was measured for the majority of clusters, that could be explained by their rather early version of the setup, as well as rather high temperature of the experiment. Even then, the estimated value for V$_{9}$ is not that far away from what we measured. In contrast, the existing theoretical studies~\cite{Liu1991,Dorantes1993,Lee1993,Ray1999,Zhao1995,Alvarado1994} are not even close to our experimental values, as they did not even show the even/odd alternation. In table~\ref{tabaim1} the calculated values are shown for comparison.
	
	\begin{table}[!ht]
		\begin{center}
			\caption{Magnetic moments predicted by references\cite{Liu1991,Dorantes1993,Lee1993,Ray1999,Zhao1995}. Note that in Ref.~\onlinecite{Dorantes1993} values depend on the exchange integral and also on the arrangement, AFM or FM. Ref.~\onlinecite{Ray1999} showed the two lowest in energy magnetic configurations. }
			\begin{ruledtabular}
				\begin{tabular}{ccccccccc}
					(n) & \onlinecite{Liu1991} & \onlinecite{Dorantes1993}$_{FM}$ & \onlinecite{Dorantes1993}$_{AFM}$ & \onlinecite{Lee1993} & \onlinecite{Ray1999}$_{a}$ & \onlinecite{Ray1999}$_{b}$ & \onlinecite{Zhao1995} & This work\\ \hline
					2 &   -    &  -    & - &  -  & -  & - & 2 & -\\
					3 &   -    &  -    & - &  -  & 3  & 5 & 1.8 & 1.2 \\
					4 &   -    &   -   & - &  -  & 0  & 0 & 1.4 & 0.01 \\
					5 &   -    &   -   & - &  -  & 1  & 3 & 0.5 & 0.78\\
					6 &   -    &   -   & -&  -  & 2  & 0 & 3 & 0.01\\
					7 &   -    &   -   & - &  -  & 0.98  & 3.1 & 0.7 & 0.86 \\
					8 &   -    &   -   & - &  -  & 2  & 2 & 1.6 & 0.004\\
					9 &  2.89 &  0-4    & 0-3 & 3  & 3  & 0.99 & 0.9 & 0.83\\
					10 &  -     &   -   & - &  -  & -  & - & 5 & 0.003\\
					11 &  -     &   -   & - &  -  &  - & - & 0.7 & 0.91\\
					15 & 0      &   0-4   & 0-4 &  1  & -  & - & 0.6 & 2.4\\
				\end{tabular}
				\label{tabaim1}
			\end{ruledtabular}
		\end{center}
	\end{table}
	
	\subsection{Niobium}
	Niobium is a 4d metal, with electronic configuration [Kr]4$d^4$5$s^1$. It is a well-known superconducting material in the bulk form. Clusters of Nb were studied before\cite{Moro2003,deHeer2004}, showing the odd-even variations in both magnetic and electric deflections. The observed unusual behavior of electric polarizability and dipole moments was interpreted as an onset of superconductivity\cite{Moro2003,deHeer2004}. Note however that the magnetic deflection profiles measured in this work are strongly different from those of Refs.~\onlinecite{deHeer2004}, showing, similar to vanadium, a perfect atom-like splitting for most of odd-numbered clusters, as demonstrated for Nb$_{11}$ cluster in Fig.\ 4(b). From all the measured cluster sizes, only Nb$_7$ and Nb$_{15}$ among the odd-numbered clusters, showed a single-sided deflection with the corresponding magnetic moments of about 3~$\mu_B$. For all the other odd-numbered clusters, the magnetic moment of 1~$\mu_B$ was measured within the error bar, all of them showing an atomic-like behavior and thus the absence of relaxation. Among the even-numbered clusters, no net magnetic moments was observed, except for Nb$_{22}$ and Nb$_{28}$, whose magnetic moment was found to be slightly larger than 1~$\mu_B$.
	
	In the work of Moro et al.~\onlinecite{deHeer2004}, the magnetic moments of 1~$\mu_B$ were derived from the broadening of the deflection profiles for \emph{all} odd-numbered clusters. The profiles for Nb$_7$ and Nb$_{15}$ are not actually shown, which makes it impossible to speculate about the reason for the difference in magnetic moments as compared to our data. The only thing mentioned is that the magnetic moment of the Nb$_7$ cluster, determined from the broadening in Ref.~\onlinecite{deHeer2004}, appears to be noticeably reduced. This may thus indicate the internal relaxation as found in our measurements. 
	
	We should note, in addition, that changing the source pressure as was done with vanadium clusters, in this case had less pronounced effect on the deflection profile, indicating a better thermalization of Nb cluster beam as compared to V. 
	
	\begin{figure}[tb]
		\centering
		\includegraphics[width=0.9\linewidth]{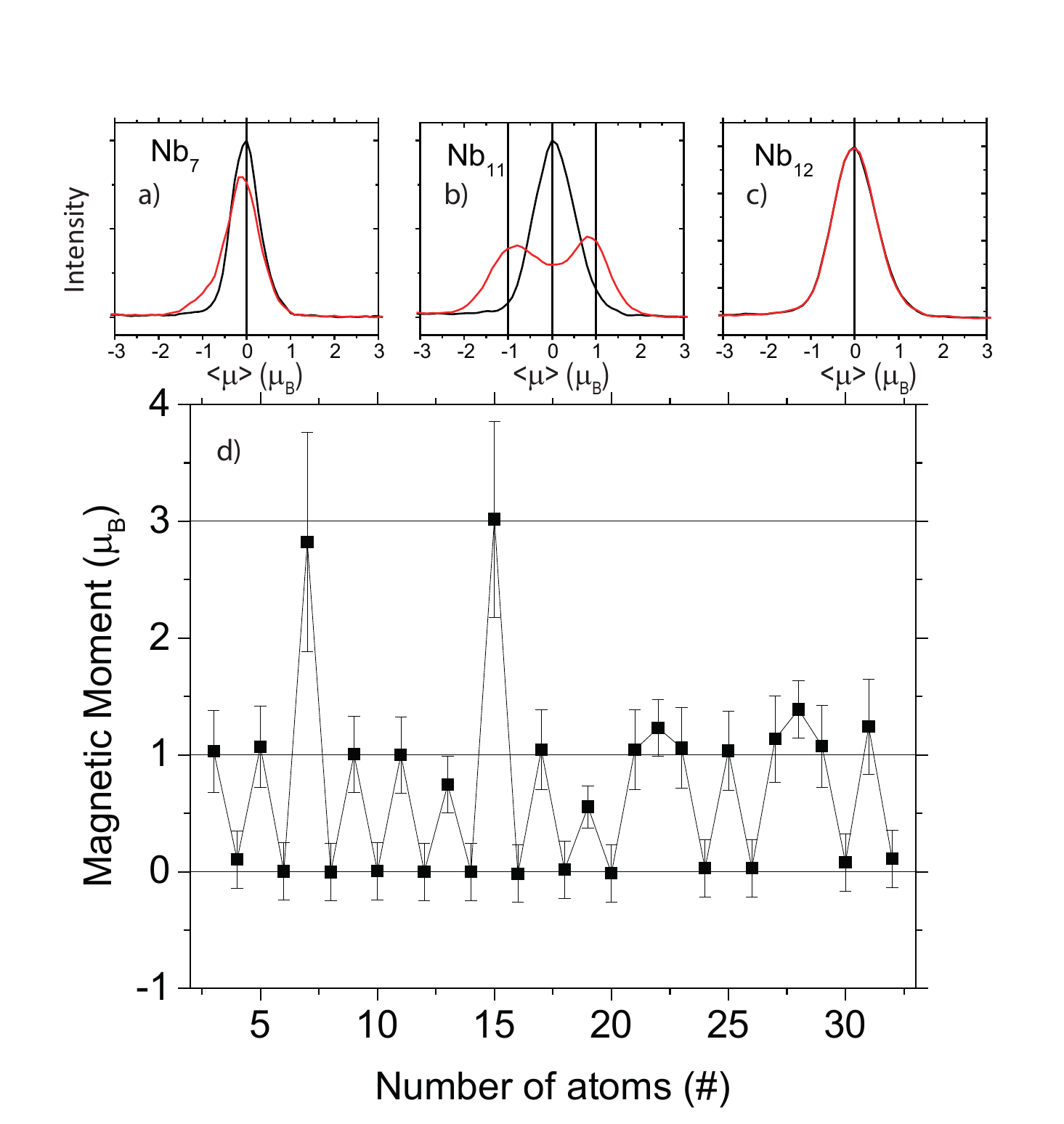}
		\caption{Top: Different kinds of behaviour for niobium clusters: a) Nb$_{7}$ cluster shows superparamagnetic behaviour; b) Nb$_{11}$ cluster that shows an atomic-like splitting in two states; c) Nb$_{12}$, which is not magnetic at all; all of them are obtained at a temperature of 25~K; black line is for zero magnetic field, while the red line corresponds to the magnetic field of 2.4~T; the lines point at $\pm$1~$\mu_B$ and 0~$\mu_B$, in case that the behavior is atomic-like. Bottom: d) Evolution of the magnetic moment as function of cluster size.}
		\label{mag-Nb}
	\end{figure}
	
	\subsection{Tantalum}
	The last kind of clusters studied in this work are tantalum clusters. Tantalum is a 5d metal, which has an electronic configuration [Xe]4$f^{14}$5$d^3$6$s^2$ and it is significantly heavier than vanadium and niobium, which makes it more challenging for precise deflection measurements. Pure tantalum clusters magnetism has not been studied experimentally so far, at least to our knowledge, only in Ref.\ \onlinecite{deHeer2004} it was briefly mentioned that their magnetic properties were similar to those of the other elements from the group V. Fa et al. \cite{Dong2006} studied these systems theoretically and determined that the magnetic moment for an even number of tantalum atoms is zero, except for Ta$_{2}$, magnetic moment of which was estimated to be 4~$\mu_{B}$, while an odd number of atoms in the cluster leads to a magnetic moment of 1~$\mu_B$. 
	
\begin{figure}[tb]
		\centering
		\includegraphics[width=0.9\linewidth]{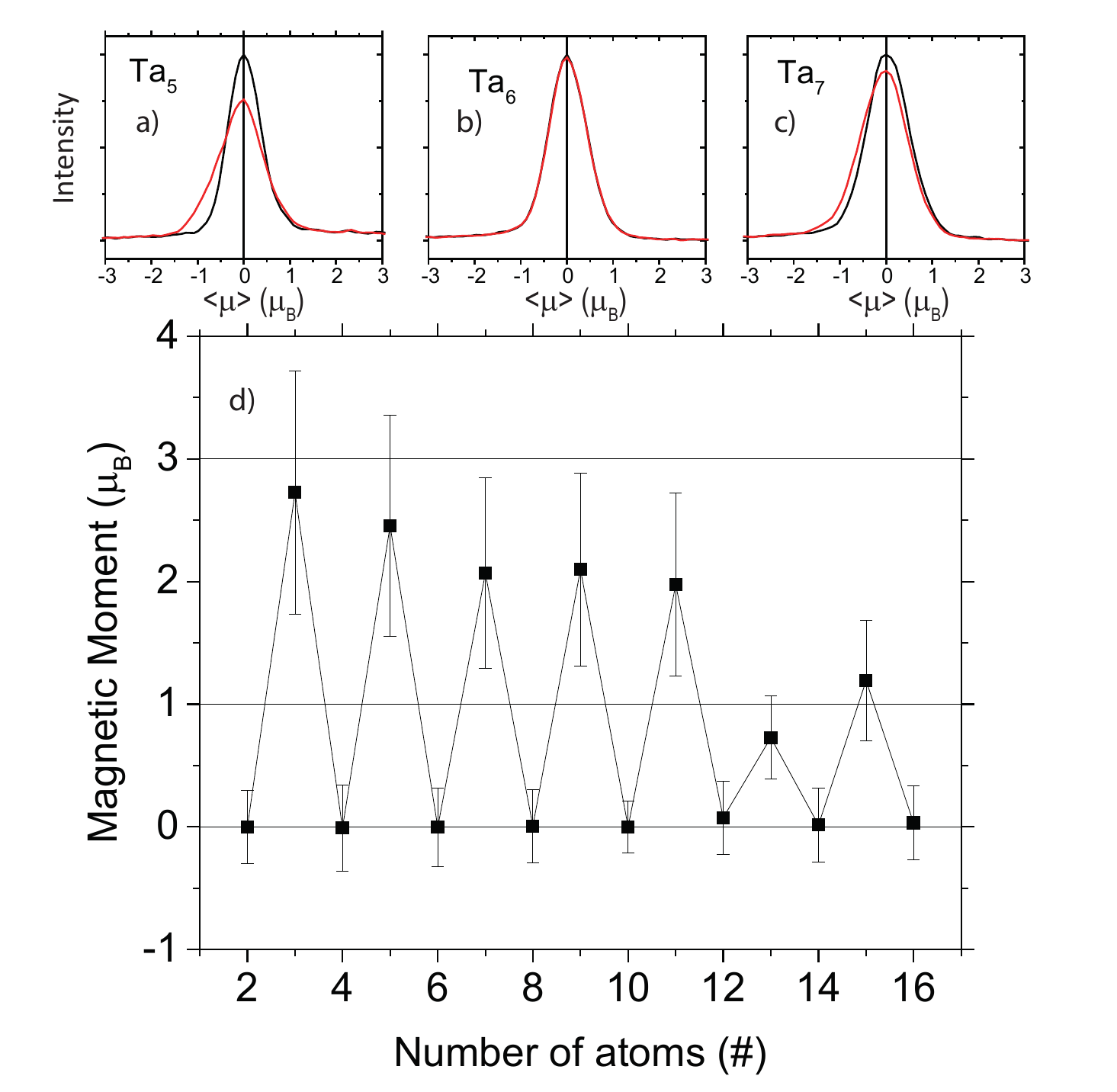}
		\caption{Top: Characteristic types of magnetic deflection for tantalum clusters: a) Ta$_{5}$, which is a superparamagnetic cluster; b) Ta$_{6}$, which is not magnetic; c) Ta$_{7}$, which again shows superparamagnetic behaviour; all of them are obtained at a temperature of 25~K; black line is for zero magnetic field, while the red line corresponds to the magnetic field of 2.4~T; the dashed lines point at $\pm$1~$\mu_B$, corresponding to the atomic-like behaviour. Bottom: d) Evolution of the magnetic moment as function of cluster size.}
		\label{mag-Ta}
	\end{figure}

	Our measurements showed again the odd-even alternation of the magnetic properties. However the main difference with the previously shown niobium and vanadium cluster is that for tantalum only single sided deflections were found, as can be seen in Fig.\ \ref{mag-Ta}. Also in Fig.\ 5(d), the evolution of the magnetic moment of tantalum clusters is shown; the magnetic moments measured range from about  3~$\mu_B$ for smaller clusters, and down to 1~$\mu_B$ for  Ta$_{13}$ and Ta$_{15}$. 
	
	The temperature dependence for tantalum is similar to the superparamagnetic clusters of vanadium and niobium, meaning that it vanishes with increasing temperature faster than it happens for atomic-like clusters.
	
	\section{Discussion}
	\subsection{Thermalization}
	As shown above, there are three different kinds of behavior for the clusters of the vanadium-group metals studied in this work. The data treatment was different depending in the behavior of every cluster, due to the nature of the peaks. 
	
	For atomic-like clusters, we did a fitting by 2 or 3 Gaussian peaks. Note that when the backing pressure was large enough, only 2 peaks were visible, while the peak at 0~$\mu_{B}$ was still present, though negligible compared to the others. The values of the magnetic moments did not depend on the amount of peaks and were the same within the error bars, showing that the relaxation in the cluster does not decrease the absolute value of the magnetic moment, but rather relaxes the spin into the lattice or not.
	
	Opposite to this, when the clusters were superparamagnetic the treatment was different; because the deflected profile is not symmetric, a Gaussian fit can not be applied. For these clusters we used the mean value of the peak instead of the maximum, as it gives a better estimation of the deflection. We should also note here that while increase of He backing pressure results in a clearly better thermalization in atomic-like clusters, the same imporvement could also be observed in the superparamagnetic ones.  
	
	We believe that this is a consequence of the decrease of vibrational modes due to the larger amount of He gas in the source, so that the vibrational modes are more relaxed due to the larger amount of collisions induced by this pressure. The amount of He is therefore a key parameter to achieve a better thermalisation, and thus also assures a better estimation of the net magnetic moment.
	
	\subsection{Kramers states and relaxation} 
	Usually, the interaction of magnetic moments with crystallographic lattice is described in terms of magnetic anisotropy. However, for magnets with the total spin S = 1/2 only intersite magnetic anisotropy is allowed by symmetry, and for a single quantum particle with S = 1/2 the anisotropy is absent completely; only Zeeman term determines spatial orientation of the spin moment at the equilibrium, as discussed in the previous section on thermalization. At the same time, kinetics of the magnetic moment is important, while a certain amount of spin-lattice relaxation still present. Generally speaking, one can expect that, due to spin-lattice relaxation, angular momentum will be transferred back and forth to rotational degrees of freedom of the cluster as a whole, and the initial state spin-up $\ket{\uparrow}$ will be transformed into a superposition $\alpha\ket{\uparrow}\ket{0}+\beta\ket{\downarrow}\ket{1}$ where $\ket{0}$ is the ground rotational state of the cluster and $\ket{1}$ is an excited state with z-projection of rotational momentum equal to one (for simplicity, we consider here as an example only the case of zero temperature). If the spin-lattice interaction energy is higher than the rotation energy quantum, one could expect many spin-relaxation events during the passage of the cluster through the gradient magnet, and therefore $\mid\alpha\mid$$\approx$$\mid\beta\mid$. In this case the average spin of the cluster will be close to zero; no deflection could be expected in such a situation. It is thus clear that the time of angular momentum transfer from the spin to the rotational degrees of freedom $\tau_{S-Lat}$ should be compared with the characteristic flight time $t_{fl}$; the condition of the atomic-like deflection reads $\tau_{S-Lat}$>$t_{fl}$. Different clusters are different by the values of $\tau_{S-Lat}$. We have to discuss therefore physical mechanisms which determine this quantity. 
	
	Spin-lattice relaxation processes (for brevity, we will call them below spin-flips) are dramatically different in the systems with integer and half-integer spins. In the latter case the Kramers theorem claims that time-reversal symmetry (which includes spin reversal) guarantees double degeneracy of all energy levels\cite{Landau1977}. This means, in particular, that for the systems with total spin $S = 1/2$, 3/2, ..., neglecting the effects of external magnetic field $H$, no static perturbation can induce the spin-flips (this is called ``Van Vleck cancellation'', see Refs.\ \onlinecite{VanVleck1940,Orbach1961}). Dynamical processes such as vibrations can break the symmetry and lead to spin reversal but not in the lowest order: either two-phonon processes or relaxation via excited states should be involved (Raman and Orbach processes, respectively\cite{VanVleck1940,Orbach1961,Abragam1970}). Both these processes are strongly suppressed at low temperatures $T$. Their probability vanishes in the limit $T\rightarrow 0$ exponentially for the Orbach processes 
	\begin{equation}
		\begin{gathered}
			1/\tau_{S-Lat}~\propto~exp^{(-\Delta/k_{B}T)},
			\label{Eq1}
		\end{gathered}
	\end{equation} 
	where $\Delta$ is the energy of the excited state involved. As for the Raman processes, in the absence of magnetic field one can write for the spin-flip probability \cite{Orbach1961} 
	\begin{equation}
		\begin{gathered}
			1/\tau_{S-Lat}~\propto~T^{9}/\Delta^{4}v^{10},
			\label{Eq2}
		\end{gathered}
	\end{equation} 
	where v is the sound velocity. In the presence of magnetic field H, the Kramers theorem is violated and additional term 1/$\tau_{S-Lat}~\propto~H^{2}T^{7}$ arises which, however, is typically much smaller than~(\ref{Eq2}) unless the Zeeman energy in this field is at least comparable to the thermal one, thus corresponding to several tens of Tesla's in our case.
	
	For non-Kramers systems (integer spin $S$) the spin-flip probability can be estimated as
	\begin{equation}
		\begin{gathered}
			1/\tau_{S-Lat}~\propto~T^{7}/\Delta^{2}v^{10},
			\label{Eq3}
		\end{gathered}
	\end{equation} 
	and is much higher than ~(\ref{Eq2}) if thermal energy is much smaller than the energy of relevant electron excitations: k$_{B}$T~$\ll$~$\Delta$. Therefore, if we are looking for the systems with anomalously long-lived total spins, we should focus our attention to the Kramers systems only. 
	
	Moreover, even within this class of systems, the case $S = 1/2$ is special, due to the absence of magnetic anisotropy. For $S > 1/2$, one can build an effective Hamiltonian of magnetic anisotropy for the ground-state multiplet, for example:
	\begin{equation}
		\begin{gathered}
			\hat{H}=K\Bigg(\hat{S}_{z}^{2}-\frac{1}{3}S(S+1)\Bigg)+E\Bigg(\hat{S}_{x}^{2}-\hat{S}_{y}^{2}\Bigg)-\mu_{B}\hat{\bm{S}}\hat{g}\textbf{H},
			\label{Hamiltonian}
		\end{gathered}
	\end{equation}
	(see, for example Ref.~\onlinecite{Ruiz2014}). 
	
	Given this, one can further consider a modulation of the anisotropy parameters $K$ and $E$ by for example atomic vibrations. However, for the case $S = 1/2$ the rigid-spin approximation used in Eq.\ (\ref{Hamiltonian}) is inapplicable in principle and multi-spin character of the Hamiltonian should be taken into account \cite{Barbara2002}. In this case, the crucial role is played by Dzialoshinskii-Moriya (DM) interactions\cite{Dzyaloshinsky1958,Moriya1960}. Formally, this is the main relativistic interaction in magnetism since it is of the first-order in spin-orbit coupling constant whereas the magnetic anisotropy is, at least, of the second-order. It vanishes however in high-symmetry systems where each of magnetic pairs has inversion center. In magnetic molecules like V$_{15}$ or Mn$_{12}$ this is typically not the case, and DM interactions play a crucial role in their magnetic properties \cite{Barbara1998,Katsnelson1999,Raedt2002,Raedt2004,Belinsky2009,Mazurenko2014}, including, probably, the magnetic tunneling behavior \cite{Raedt2002,Raedt2004}. Importantly, DM interactions do not conserve the total spin initiating transitions between the states with different multiplicity (in the lowest order, $S \rightarrow S \pm 1$). The spin-flip processes in this case involve virtually excited states with S > 1/2, each of elementary step of such transitions involve a small parameter $D/J$ where $D$ and $J$ are characteristic values of DM and exchange interactions, respectively. For magnetic clusters with $S = 1/2$ DM interaction should be the main factor responsible for the spin flips, similar to the case of V$_{15}$\cite{Barbara2002,Raedt2004}. The low probability of the spin flips observed in our experiments is consistent with the model that it is not an antiferromagnetic arrangement of several ordered spins, but indeed a single spin of $S=1/2$, that is responsible for the observed magnetic deflection. 
	
	For $S = 3/2$, in contrast, additional channels of magnetic relaxation arise due to possibility of anisotropy-induced transitions within the ground state multiplet, according to the Hamiltonian (4). 
	
	Ta is heavier than V and Nb which has two important consequences. First, the spin-orbit coupling and therefore the value of DM interactions should be much higher in Ta clusters than in those of V and Nb. Second, the phonon frequencies are considerably lower because of large atomic mass, which should essentially increase the probability of the Raman relaxation processes: their probability according 
	to Eq.\ (2) is proportional to 1/v$^{10}$ $\propto$ M$^{5}$, where M is the nuclear mass. It is impossible to say without quite cumbersome calculations which factor is more important but, anyway, they both work in the same direction.
	
	\section{Conclusions}
	Our experiments have clearly demonstrated the presence of Kramers blocking of spin-lattice relaxation in clusters of vanadium group elements that are as large as V$_{21}$. This blocking can be lifted by either Raman relaxation via an excited vibrational state, or by Orbach mechanism in clusters with larger magnetic moments. The relaxation is clearly more efficient in clusters with larger nuclear mass such as Ta. These results thus demonstrate that gas-phase clusters represent an ideal model system to study quantum coherent phenomena at practically macroscopic scales. 
	
	\section{Acknowledgments}
	We gratefully acknowledge the Nederlandse Organisatie voor Wetenschappelijk Onderzoek (NWO-I) for their financial contribution. M.I.K. acknowledges a support by European ResearchCouncil (ERC) Grant No. 338957. Moreover we would like to thank Remko Logemann, Lars Peters, Valeriy Chernyy and Joost Bakker for fruitful discussions and Erwin Muskens for technical assistance.

\end{document}